%% file: paper.tex
\documentclass[12pt]{article}
\usepackage{epsfig}

\input econfmacros.tex

\def\tom{\tilde{\omega}}
\def\toml{\tilde{\omega}_{\rm L}}

\def\rl{R_{\rm L}}
\def\bp{B_{\rm P}}

\def\up{u_{\rm P}}
\def\vp{v_{\rm p}}
\def\msun{{\rm M}_{\odot}}
\def\omf{\Omega_{\rm F}}

\def\rg{r_{\rm g}}

\def\sigm{\sigma}

\def\Title#1{\begin{center} {\Large {\bf #1} } \end{center}}

\begin{document}

\Title{Relativistic MHD jets and the GRBs}

\bigskip\bigskip


\begin{raggedright}  

{\it Christian Fendt\index{Fendt, Ch.}\\
Institut f\"ur Physik, Universit\"at Potsdam\\
Am Neuen Palais 10\\
D-14469 Potsdam, GERMANY}
\bigskip\bigskip
\end{raggedright}

\section{Introduction -- astrophysical jets}
High velocity highly collimated beams of plasma -- the jets --
are known as a general phenomenon among astrophysical sources of 
different energy and spatial scales.
At the upper end of the scale are the extragalactic jets emanating
from active galactic nuclei. 
Little brothers of extragalactic jets has been detected recently
as Galactic superluminal sources -- the so-called microquasars 
\cite{mira94}.
These jets originate in stellar mass black hole accretion disks in
high-energy binary systems.
Protostellar jets are non-relativistic with
velocities of several 100\,km/s \cite{mund90}.

The idea of a common astrophysical jet formation scenario is motivated by the
observational fact that jet formation is always connected to 
(i) the existence of (strong) {\em magnetic fields} and 
(ii) the presence of an {\em accretion disk}.
The magnetohydrodynamic (MHD) model of jet formation 
(cf.~\cite{blan82,came86})
allows to describe the various scales of astrophysical jets 
by a scaling of the jet {\em magnetization}.
The theory tells us that highly relativistic jets
must be also highly magnetized \cite{fend96} and that
rotating MHD flows are subject of a {\em self-collimation}
process \cite{heyv89, ouye97}.
In this paper, we ask whether the ultra-relativistic motion and 
strong collimation indicated for GRBs is achieved by a mechanism
similar to the ``classical'' astrophysical jets.

\section{Magnetohydrodynamic jets}
%
%
Highly relativistic MHD jets originate in the accretion disk surrounding 
a central black hole (Fig.\,1).
The disk provides the Poynting flux and drives the electric
current system.
The jet is initiated as a slow wind from the inner disk by
a process which is not yet completely understood,
in particular its time-dependent character.
Most probably, some disk instability is responsible for the ejection
of knots perpendicular to the disk surface.
The disk wind is first launched magneto-centrifugally \cite{blan82}
and further accelerated and collimated into a narrow beam by
Lorentz forces.
The parallel component 
$\vec{F}_{L,||} \sim \vec{j}_{\perp} \times \vec{B}_{\phi}$
is accelerating (or decelerating), while
the perpendicular force
$\vec{F}_{L,\perp} \sim \vec{j}_{||} \times \vec{B}$
is the collimating component (Fig.\,1).
Theoretical modeling, therefore, requires to solve the governing
MHD equations.
In this contribution, we will concentrate on solutions to the
{\em stationary, axisymmetric, ideal MHD}
equations in the relativistic limit.
The advantage of the stationary model is the possibility to obtain 
{\em global} solutions for the jet structure on spatial scales and
resolution which cannot (yet) be reached by time-dependent simulations
\cite{uchi85, koid98}.
Long-term MHD simulations can, however, demonstrate the
self-collimation of MHD jets (Fig.2; \cite{ouye97}).

\begin{figure}
\centering
\includegraphics[width=7cm]{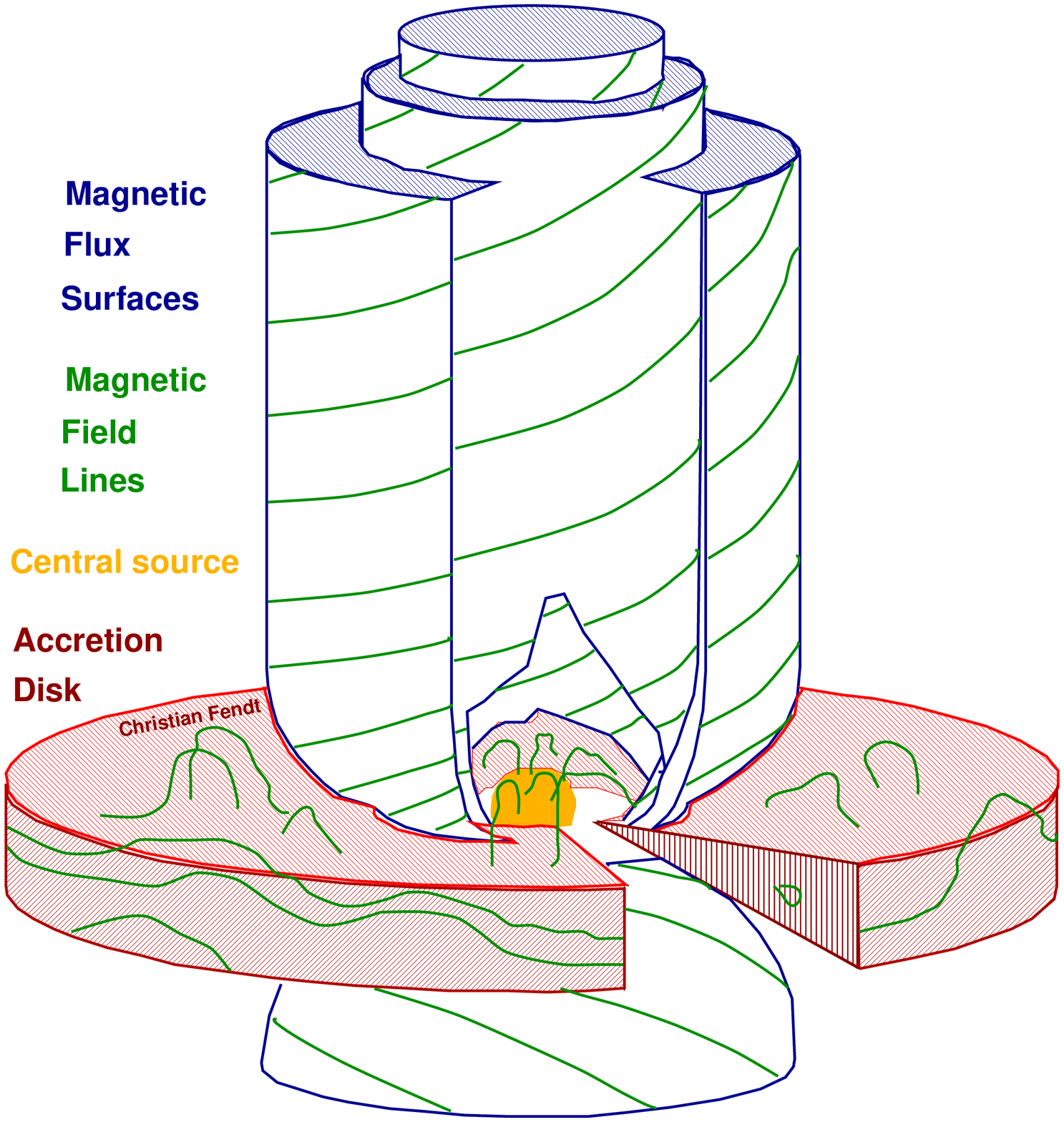}
\includegraphics[bb= 0 -50 635 327, width=7cm]{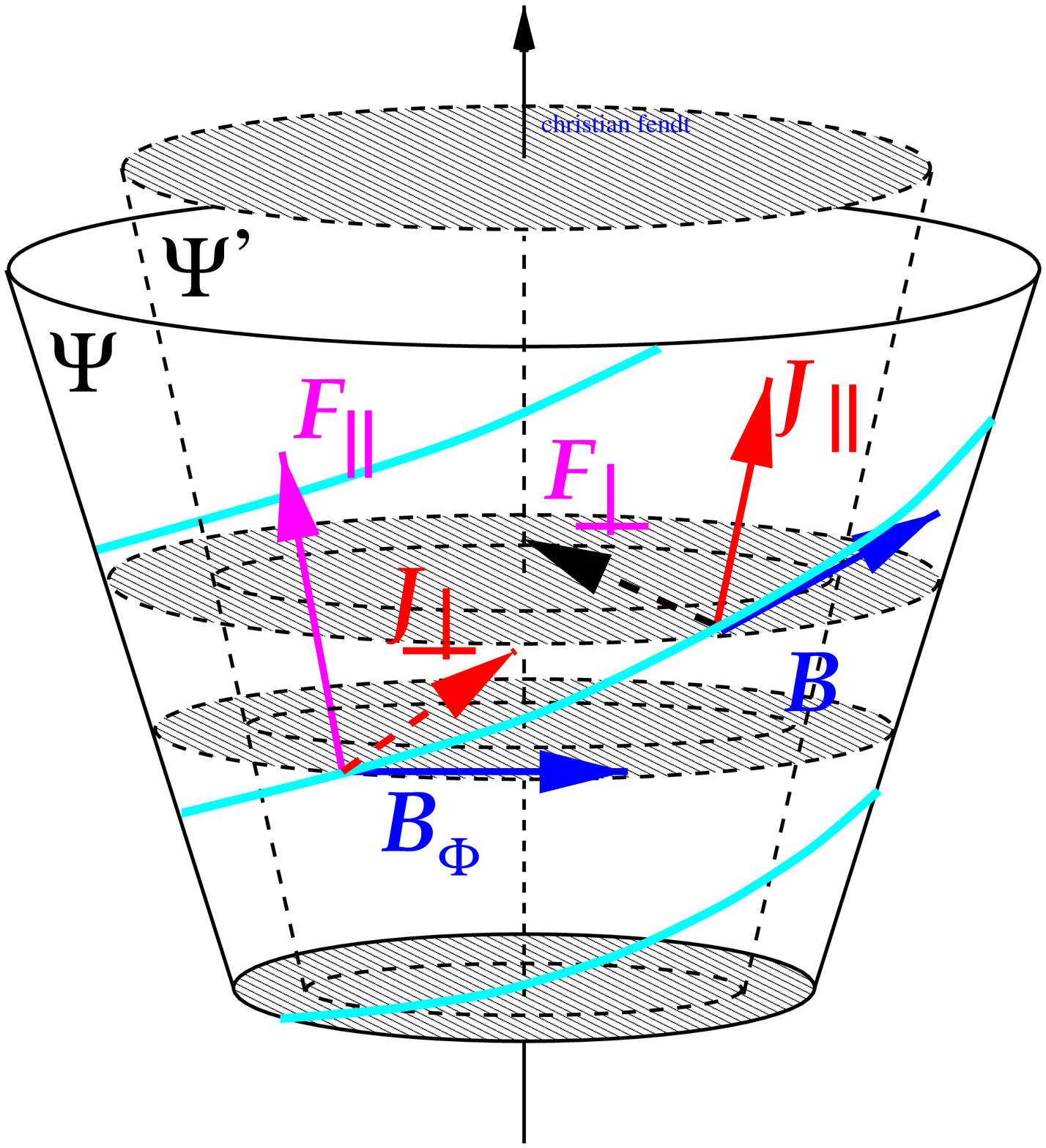}
\caption{({\it Left}) The model of MHD jets launched from an accretion disk surrounding
the central mass.
({\em Right}) The accelerating and collimating Lorentz force components are projected 
with respect to the magnetic flux surfaces $\Psi(R,Z)$.
}
\end{figure}

\begin{figure}
\centering
\includegraphics[width=13.3cm]{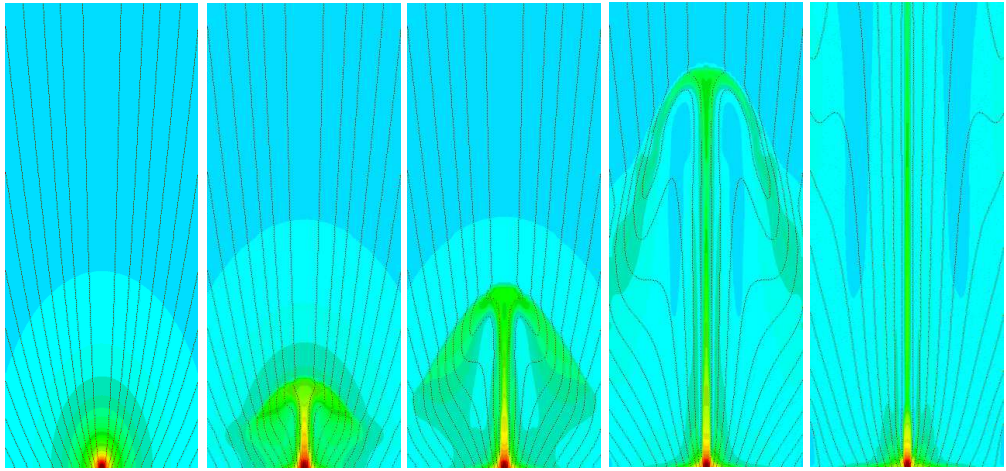}
\caption{Non-relativistic MHD simulations of jet formation. 
Density (colors) and poloidal 
magnetic field lines after 0, 50, 100, 200 and 400 inner disk rotations.
}
\end{figure}

\subsection{Relativistic stationary MHD}
The axisymmetric stationary ideal MHD equations can be reduced into
two governing equations --
the {\em Grad-Shafranov} equation (GSE) describing the force-balance
across the magnetic flux surfaces\footnote{
The magnetic flux is 
$
\Psi (r,\theta) = (1/2\pi) \int {\vec {B}}_{\rm P} \cdot d{\vec{A}}.
$
In general relativity, the integration is over a loop of the Killing
vector $\vec{m} = \tom^2\nabla\phi$.
In Minkowski space the flux is integrated over a circular area around
the symmetry axis.
}
$\Psi(r,\theta)$ and
the {\em wind equation} (WE) considering the forces along the
field.
Both equations are interrelated, however, in the relativistic limit of
a high magnetization the field structure is almost force-free and 
can be investigated independently of the flow dynamics.
Here, we will discuss solutions of the force-free GSE in Kerr metric
appropriate for jets from rotating black holes,
\begin{equation}
\tom \nabla \cdot
\left( {\alpha\,\frac{1 - \left({\tom}/{\tom_{\rm L}}\right)^2}{\tom^2}}\nabla\Psi\right)
= \tom\,\frac{\omega-\omf(\Psi)}{\alpha c^2}\,|\nabla\Psi|^2 
  \frac{d\omf(\Psi)}{d\Psi}
- \frac{1}{\alpha\tom}\frac{2}{c^2}\;I(\Psi) \frac{dI(\Psi)}{d\Psi}
\end{equation}
(see \cite{okam92}).
In Eq.\,(1), $\omega $ is the angular velocity of a zero angular momentum 
observer (ZAMO), 
$\alpha $ the lapse of ZAMO proper time to the global time,
$\tom$ the cylindrical radius, 
and $\toml(\Psi)$ the position of the two light ``cylinders'' (l.c.).
$\omf(\Psi)$ can be considered as rotational velocity of magnetic field lines.
$I(\Psi)$ is the total poloidal electric current.
The general relativistic {\em wind equation} 
for the poloidal velocity
$\up \equiv \Gamma \vp/c$
is\footnote{
with the following abbreviations,
$
k_0 \equiv g_{33} \omf^2 + 2 g_{03} \omf + g_{00},
$
$
k_2 \equiv  1 - \omf (L/E),
$ and
$
k_4 \equiv
- \left(g_{33} + 2 g_{03} (L/E) + g_{00}(L/E)^2 \right) /
         \left( g_{03}^2 - g_{00} g_{33} \right)
$
}
\cite{came86,taka90,fend01}
\begin{equation}
u_{\rm p}^2 + 1  = \left(\frac{E}{\mu}\right)^2
\frac {k_0 k_2  - 2 k_2 M^2 - k_4 M^4}{(k_0 - M^2)^2},
\end{equation}
with the Alfv\'en Mach number $M$.
Specific total energy and angular momentum density $E(\Psi)$, 
$L(\Psi)$, are conserved quantities along $\Psi$.
As a key parameter for MHD jets, 
the magnetization $\sigm(\Psi)$ measures the Poynting flux
in terms of mass flux,
\begin{equation}
\sigm = \frac{\Phi^2\omf^2}{4 \dot{M} c^3},
\label{eq_sigdef}
\end{equation}
with the magnetic flux function $\Phi = \bp R^2$ and the mass flux
$\dot{M}$.
The launching of a relativistic jet requires at least one of three 
conditions -- rapid rotation, strong field, or low mass load (Eq.\,3).
In general, {\em relativistic jets are highly magnetized}
\cite{mich69,came86,fend96}.
Fixing the field distribution, one finds a relation between the
asymptotic Lorentz factor $\Gamma_{\infty}$ and $\sigm$.
We have (Michel-scaling)
$
\Gamma_{\infty} = 
\sigm^{1/3} \simeq
(\omf^2\,\Phi^2/\dot{M}_{\rm jet})^{1/3}
$
for a spherical flow with negligible gas pressure (``cold'') 
\cite{mich69}. 
In the case of {\em collimation}, the power law index is
different.
If the magnetic flux surfaces open up faster than spherically
radial, 
the asymptotic {\em flow is dominated by kinetic energy}
\cite{bege94}.

\begin{figure}
\includegraphics[bb= 18 244 592 518,width=8cm]{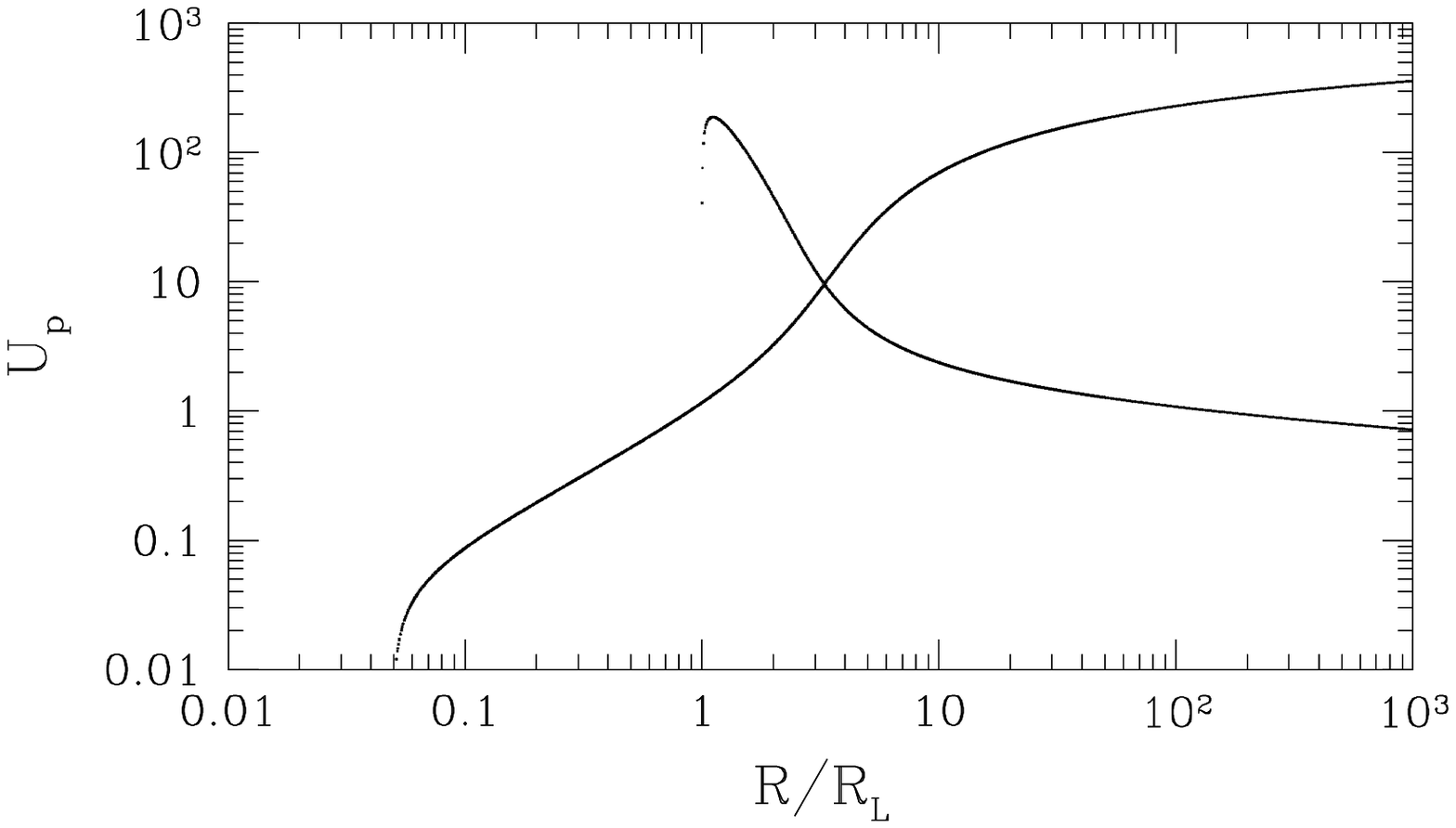}
\includegraphics[bb= 18 244 592 518,width=8cm]{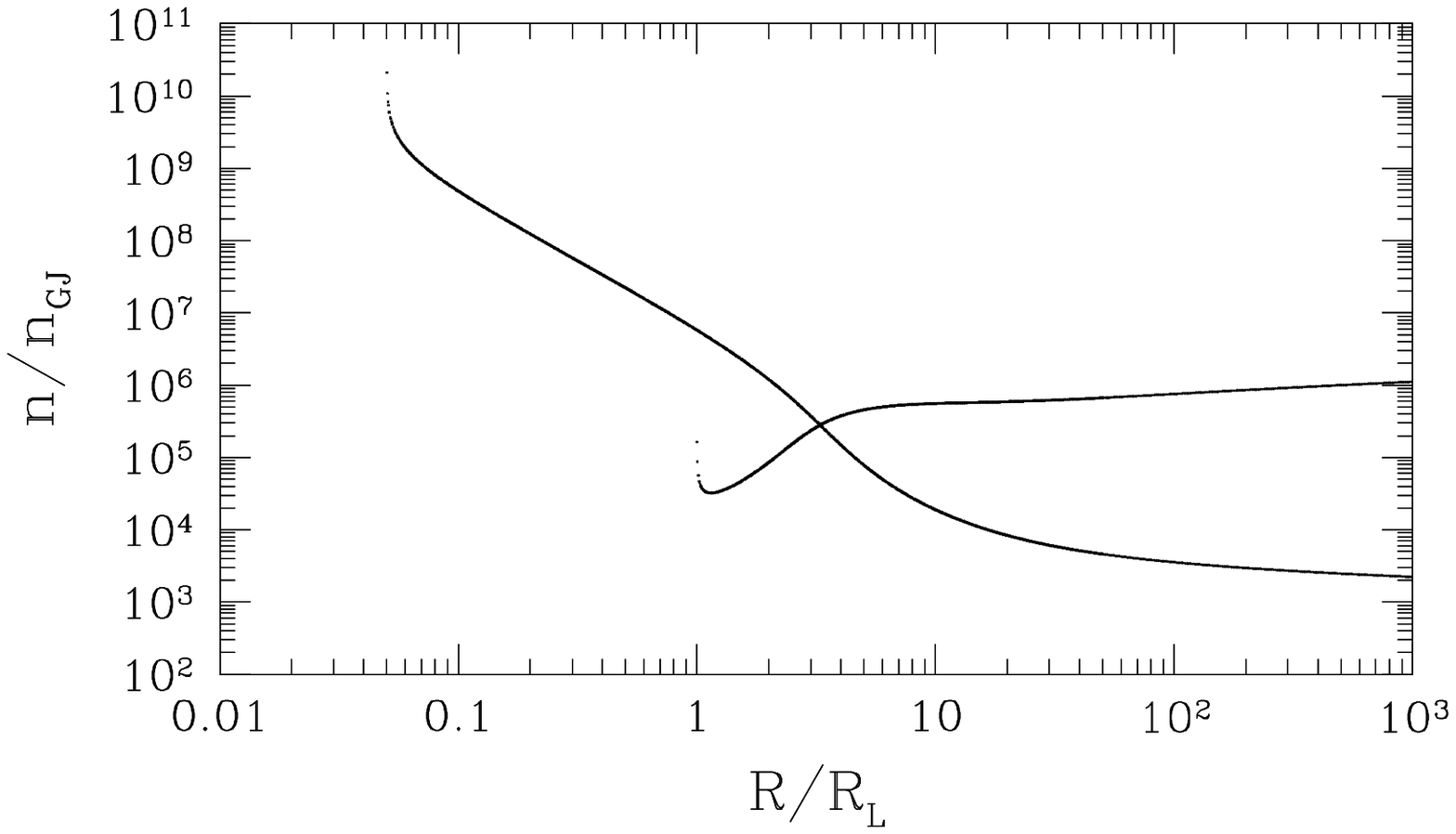}
 \caption{
Solution of the relativistic MHD wind equation.
Poloidal velocity $\up$ and particle density $n$ over Goldreich-Julian density
$n_{\rm GJ}$ versus radius.
At the fast magnetosonic point, $R=33\rl$, both solution branches intersect.
}
\end{figure}

\section{MHD jets in GRBs -- structure and dynamics}

\subsection{Parameter estimates}
We now estimate MHD jet parameters in the framework of GRBs
assuming that these jets originate close to
a stellar mass black hole and reach Lorentz factors up to 1000.

\smallskip
{\bf Disk magnetic field:}
Differential rotation amplifies the toroidal field
up to equi\-par\-ti\-tion, $B_{\phi} < B_{\rm eq}$.
A disk dynamo may amplify poloidal fields up to
$B_{\rm P} < B_{\phi}$.
In ad\-vec\-tion dominated disks, 
$
B_{\rm eq} \simeq 10^9 {\rm G}\,(M/5\msun)^{-1/2}
(\dot{M}_{\rm a}/\dot{M}_{\rm E})^{1/2}
(R/3\rg)^{-5/4},
$
with central mass $M$, accretion rate $\dot{M}_{\rm a}$ 
in terms of Eddington $\dot{M}_{\rm E}$, 
and the horizon at $\rg$
\cite{nara95}.
GRB models considering $10^{15}$G fields are not consistent
with this estimate.

\smallskip
{\bf Light cylinder:}
The l.c. of the jet magnetosphere is located at 
$\rl \simeq c/\omf\simeq 4\times 10^6{\rm cm}\left(M/5\msun\right)$
if the jet originates just outside the marginally stable orbit.

\smallskip
{\bf Magnetization:}
The upper limit magnetization can be derived assuming the jet
origin at the marginally stable orbit 
(highest field strength, most rapid disk rotation).
Jet mass flow rates $\sim 10^{-11}\msun\,{\rm yr}^{-1}$
are indicated by GRB afterglows constraining the baryonic 
``jet'' mass to $10^{-4}\msun$.
This gives a jet magnetization of about $10^4$.

\subsection{Acceleration -- asymptotic Lorentz factor}

Figure\,2 shows a solution of the cold relativistic MHD WE
in the Minkowski limit\footnote{
For hot wind solutions considering the influence of Kerr metric on the 
jet acceleration, see \cite{fend01}}
($\sigm=1000$, $Z(R) \sim R^{1.8}$).
The flow accelerates rapidly, becomes super Alfv\'enic ($R>\rl$),
super fast magnetosonic ($R>33\rl$)
and reaches asymptotically $\up \sim 300$.
The density remains above the Goldreich-Julian density, 
confirming the underlying {\em MHD assumption} a posteriori.
For cold jets, $\sigm$ is a free parameter.
Thus, we may investigate 
the jet dynamics for different magnetization (Fig.\,3).
The two chosen flux distributions deviate from a spherical flow
and indicate the modified Michel-scaling.

\begin{figure}
\begin{minipage}{7cm}
\includegraphics[bb= 18 244 592 518,width=8cm]{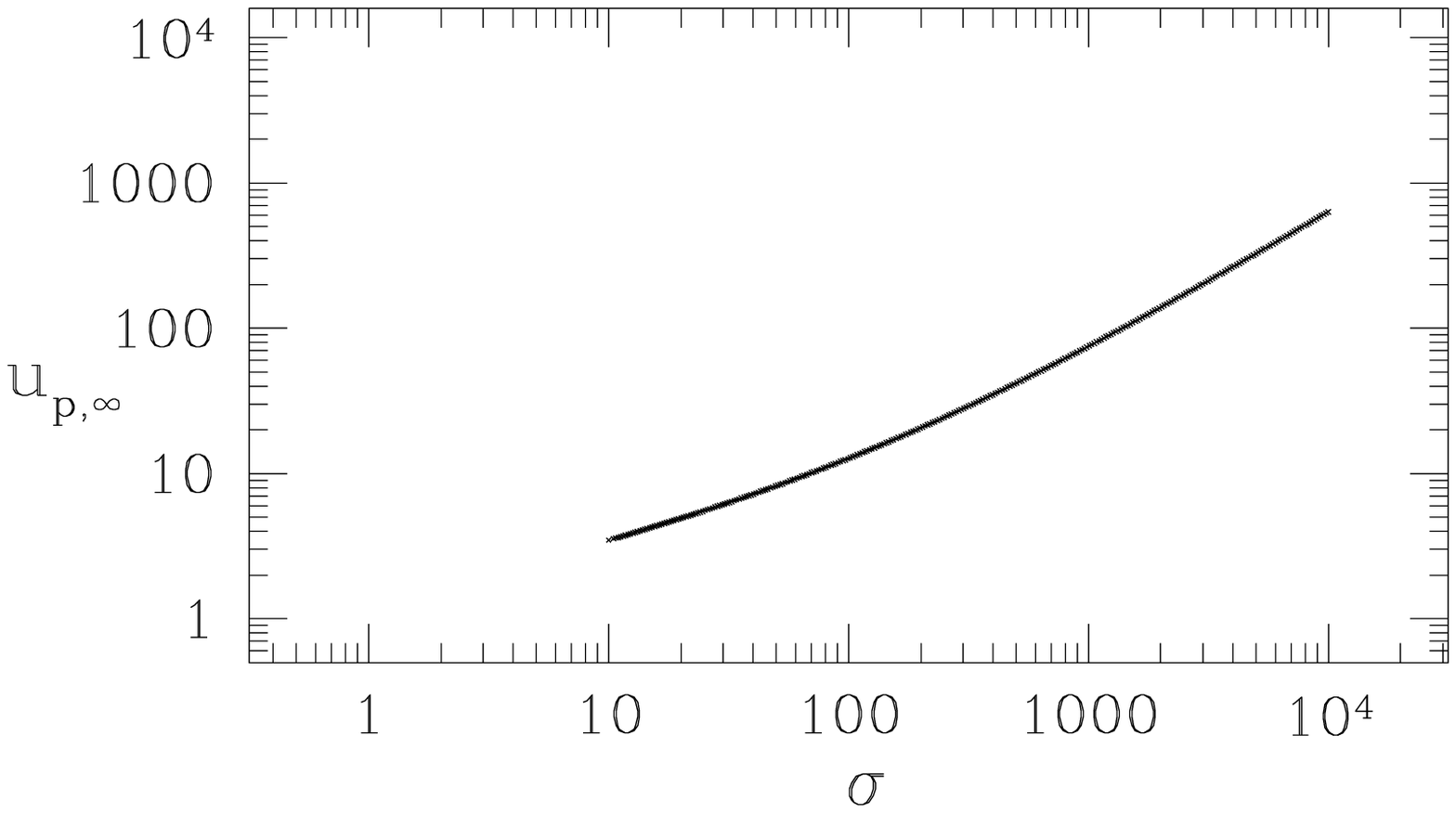}
\includegraphics[bb= 18 244 592 518,width=8cm]{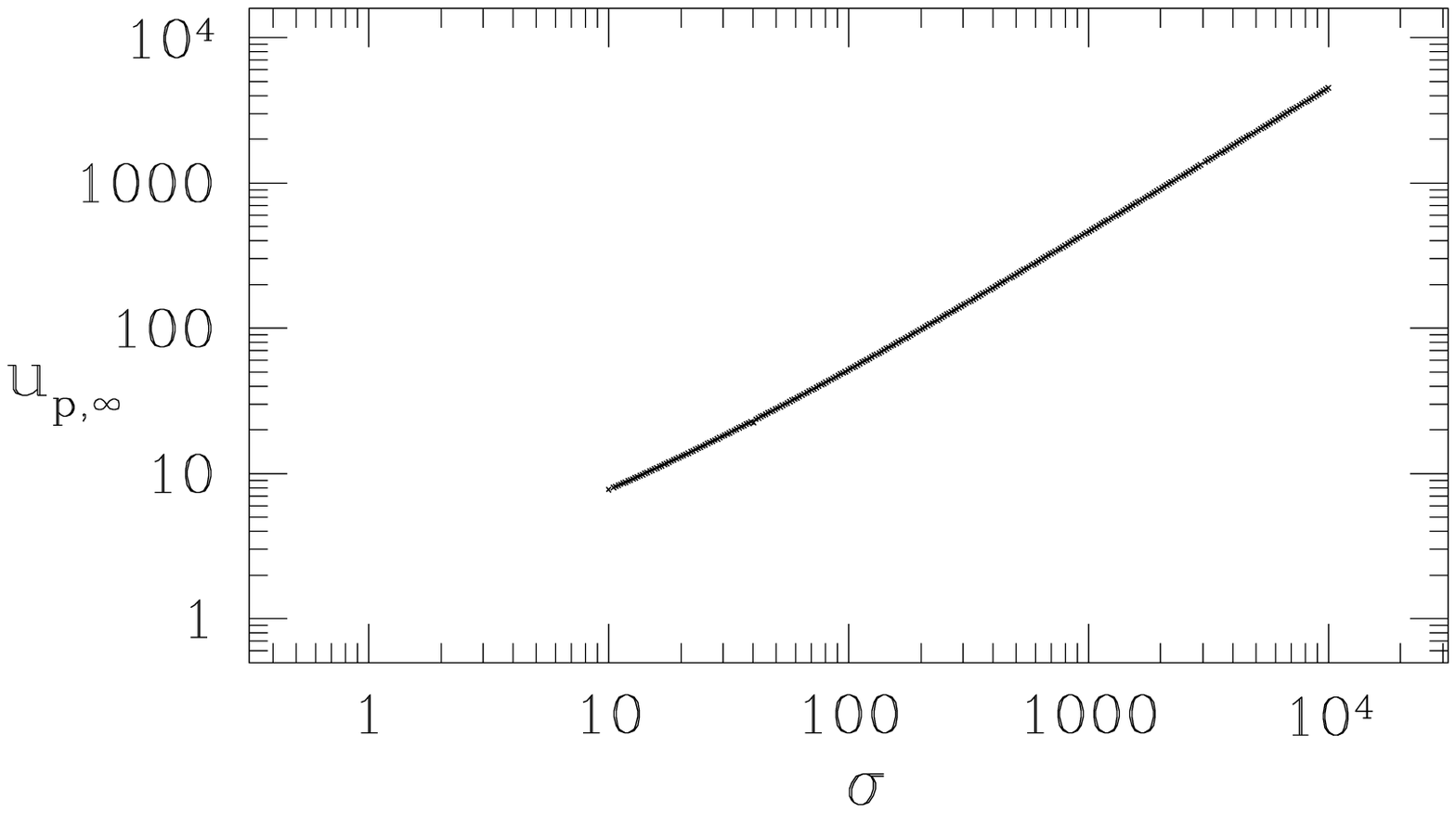}
\end{minipage}
\begin{minipage}{8cm}
\includegraphics[width=8cm]{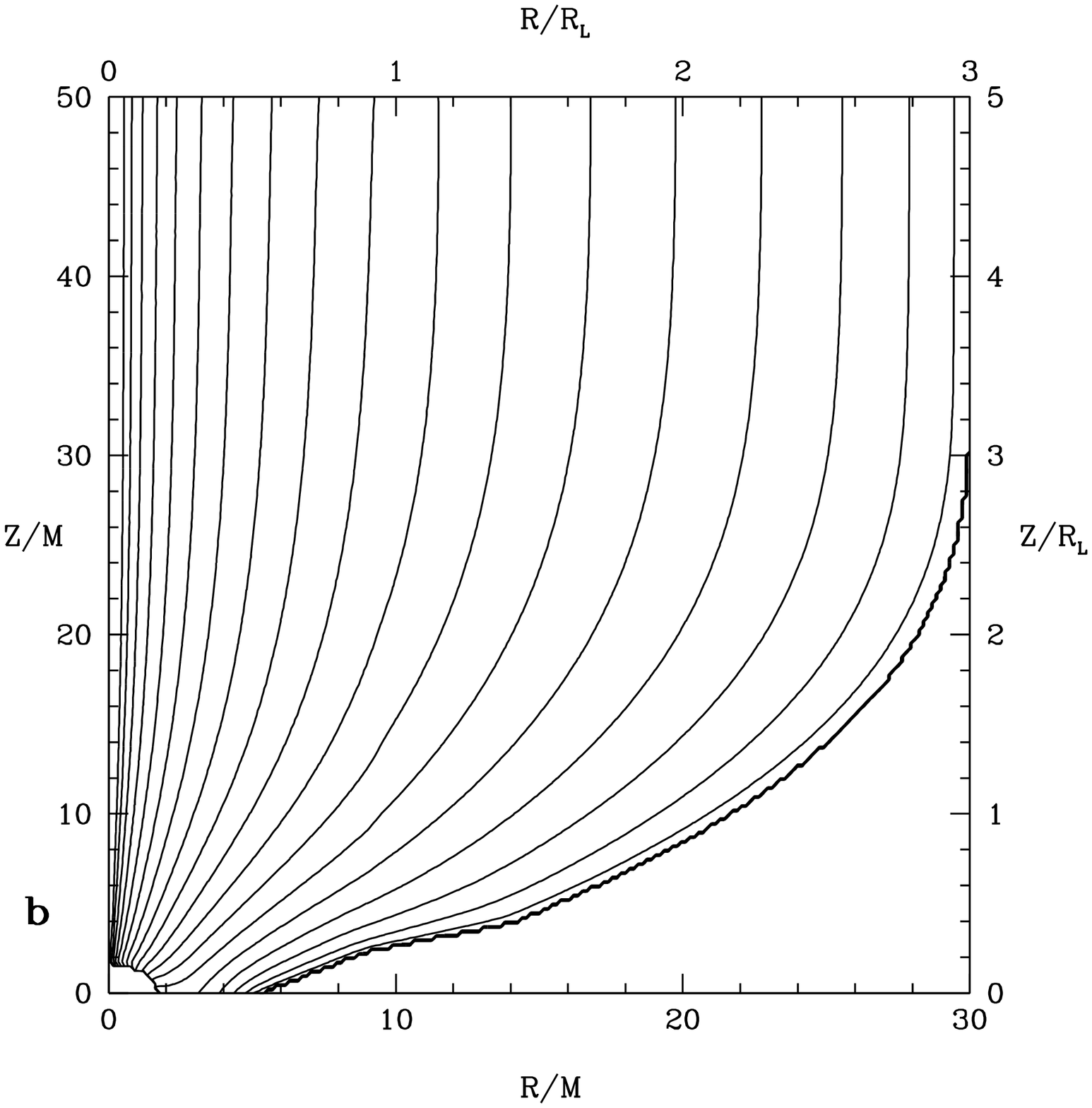}
\end{minipage}

\caption{
({\it Left}) Modified Michel-scaling, $u_{\rm p,\infty } (\sigm)$ 
for $B_pR^2\!\sim\!R^{-0.01}$({\it top}) and $\sim\!R^{-0.1}$({\it bottom}).
({\it Right}) Global magnetic field structure of the collimation jet $\Psi(R,Z)$.
}
\end{figure}

\subsection{Collimation}
Figure\,3 shows a solution of the force-free GSE extending
from the inner l.c. close to the hole,
across the outer l.c. to the asymptotic, cylindrically 
collimated jet
($a=0.8$, $\Psi_{\rm max} = 10^{21}{\rm G\,cm^2}$,
$I_{\rm max} = 10^{15} {\rm A}$, $\rl=10^7{\rm cm}$).
The shape of the collimating jet is determined by the 
l.c. regularity condition.
The field structure close to the hole and disk indicates a hollow
jet with the bulk of the mass flow in the outer jet layers 
\cite{fend97}.

\subsection{MHD jets in GRBs -- open questions}

The numerical solutions presented here show that the classical
MHD jet formation scenario can indeed be extended to GRB parameter 
in order to achieve both ultra-relativistic velocities and strong
collimation.
However, serious questions remain un-answered, among them: 
(i) Due to the Michel-scaling, it is difficult to account for a
wide range in Lorentz factors, $10^2 < \Gamma < 10^5$, which
seems to be needed to generate the temporal variability of GRBs.
(ii) Also, the bimodal distribution of GRB duration does not fit into
the single MHD jet model.
It is not clear, how such questions can be answered within a MHD jet
model or whether pure electromagnetic or plasma processes
are actually dominant.

\bigskip
\bigskip

\noindent
{\bf Acknowledgments.}
I thank NORDITA for the financial support.


%
\def\Discussion{
\setlength{\parskip}{0.3cm}\setlength{\parindent}{0.0cm}
     \bigskip\bigskip      {\Large {\bf Discussion}} \bigskip}
\def\speaker#1{{\bf #1:}\ }
\def\endDiscussion{}
\Discussion

\speaker{Blackman}
Can you comment on the claims of Ustyugova et~al.
that the boundary shape strongly influences the presence
or absence of collimation?

\speaker{Fendt} 
We did not encounter this problem, maybe due to different
boundary conditions in ZEUS code. 
Certainly, Ustyugova et~al.'s point should be carefully
considered.

\speaker{Hujeirat}
Do you predict the existence of a super-Keplerian layer above
the disk, especially as you said you do agree with the
Blandford \& Payne 1982 scenario?

\speaker{Fendt}
The sub-Alfv\'enic disk wind co-rotates with the magnetic field.
The toroidal velocity increases linearly $v_{\phi} \sim R$ along
the flow. Beyond the Alfv\'en point, $v_{\phi} \sim R^{-1}$.

\endDiscussion
 
\end{document}

%% file: econfmacros.tex
\def\beq{\begin{equation}}
\def\eeq#1{\label{#1}\end{equation}}
\def\eeqn{\end{equation}}

\def\beqa{\begin{eqnarray}}
\def\eeqa#1{\label{#1}\end{eqnarray}}
\def\eeqan{\end{eqnarray}}

\let\bar=\overbar

\def\Dslash{\not{\hbox{\kern-4pt $D$}}}
\def\dslash{\not{\hbox{\kern-2pt $\del$}}}

\def\msb{{\bar{\ssstyle M \kern -1pt S}}}

